# Self-Siphon Simulation Using Molecular Dynamics Method

SPARISOMA VIRIDI[a], SUPRIJADI[b], SITI NURUL KHOTIMAH[a], NOVITRIAN[a], FANNIA MASTERIKA[c]

[a]Nuclear Physics and Biophysics Research Division, Faculty of Mathematics and Natural Sciences, Institut Teknologi Bandung, Jl. Ganesha 10, Bandung 40132 Indonesia, E-mail: dudung@fi.itb.ac.id, nurul@fi.itb.ac.id, novit@fi.itb.ac.id

[b]Theoretical High Energy Physics and Instrumentation Research Division, Faculty of Mathematics and Natural Sciences, Institut Teknologi Bandung, Jl. Ganesha 10, Bandung 40132 Indonesia, E-mail: supri@fi.itb.ac.id

[c]Magister in Physics Teaching Program, Faculty of Mathematics and Natural Sciences, Institut Teknologi Bandung, Jl. Ganesha 10, Bandung 40132 Indonesia, E-mail: fannia_masterika@yahoo.com

**Abstract.** *A self activated siphon, which is also known as self-siphon or self-priming siphon, is simulated using molecular dynamics (MD) method in order to study its behavior, especially why it has a critical height that prevents fluid from flowing through it. The trajectory of the fluid interface with air in front of the flow or the head is also fitted the trajectory modeled by parametric equations, which is derived from geometry construction of the self-siphon. Numerical equations solved using MD method is derived from equations of motion of the head which is obtained by introducing all considered forces influencing the movement of it. Time duration needed for fluid to pass the entire tube of the self-siphon, τ, obtained from the simulation is compared quantitatively to the observation data from the previous work and it shows inverse behavior. Length of the three vertical segments are varied independently using a parameter for each segment, which are $N_5$, $N_3$, and $N_1$. Room parameters of $N_5$, $N_3$, and $N_1$ are constructed and the dependency of τ to these parameters are discussed.*

## 1 Introduction

A self-siphon is a siphon that can be primed by itself. The oldest version of this apparatus could be the one that is invented by Leonardo da Vinci [1], a cow horn-like shape siphon, as illustrated in Codex Leicester [2]. It can be used in wide range of applications, such as unique siphon with hydraulic capillary and self-venting as microfluidic tehnologies for nuleic acid extraction [3], as large siphon system to convey water over a dam or retaining wall [4], as part of wastewater treatment [5] and Hydro-Jet Screen [6], and as apparatus in stabilizing wastewater ponds [7] and treating milking parlour wastewater in cold climate [8]. Some applications do need the advantage of a self-siphon, such for subcritical supercritical flows [9]. Seeing the uses of selfsiphon, many patents based on its feature has been reported, such as a self-priming siphon filled with hydrophilic material for irrigation [10] and self-priming siphon for removing liquid from a dead storage of a elevated storage tank [11]. Many mechanisms have been used in inventing a self-siphon, which are where the flow is induced by creating a partial vacuum within priming chamber [12], where self-siphon is using one-way valve and must be inserted several times [13], where self-starting siphon has small bottle filled with air [14], or where self-siphon has several unequal length of bends [15, 16]. The last type of self-siphon will be discussed in this work. Experiments continuing previous work [17] have been conducted using segmented self-siphon and length variation for three vertical straight segments has been done [18]. Simulations are performed to explain the phenomenon using molecular dynamics method, which is simple than using two-fluid model that has already implemented for flow in a siphon [19].

## 2   Simulations

In order to make a model of the self-siphon, the real example in Figure 1 is mimiced also in several segments as illustrated in Figure 2. Three vertical straight segments, which are first, third, and fifth segment, have length that can be modified using several parameters: $N_1$, $N_3$, and $N_5$. There are total seven segments, whose details are illustrated in Figure 2. Small segment in first, third, and fifth segment has lenght of w, which is chosen to be the same as in experiments, 2 cm. Other dimension parameters are $R_0 = R_2 = R_4 = 1$ cm, $L_6 = 15$ cm, $\theta_0 = 0.75\pi$, $\theta_2 = \theta_4 = \pi$, and inner diameter of the tube $d = 0.6$ mm.

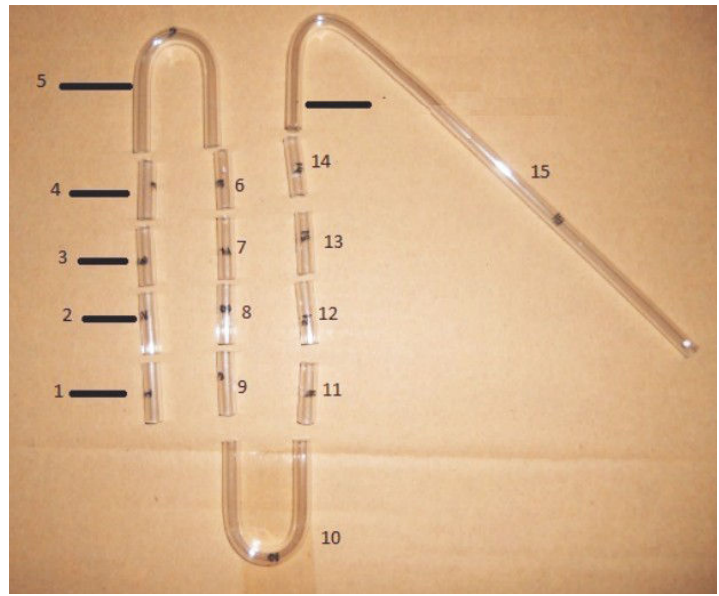

Figure 1. Experimental configuration [17] of self-siphon indicated by numbers: 15 is zeroth and sixth segment, 11-14 is first segement, 10 is second segment, 6-9 is third segment, 5 is fourth segmen, and 1-4 is fifth segment. Notice that zeroth, second, and fourth segment already contribute one small segment to first, third, and fifth segment.

As the self-siphon immersed into water, water will flow from inlet (fifth segment) to outlet (sixth segment). Time needed to accomplish all segments is defined as τ. Parameters $N_5$, $N_3$, and $N_1$ are varied and its influence on τ is observed.

Equations of motion a head, the fluid interface with air in front of the flow, with mass *m*, cross section *A*, and thickness $\Delta h$, is constructed, where forces applied on it is illustrated in Figure 3. There are two types of segment which have different equations of motion: straight segment and semi-circular path. Fifth, third, first, and sixth segments are part of the first type of segment and zeroth, second, and fourth segments are part of the later type of segment.

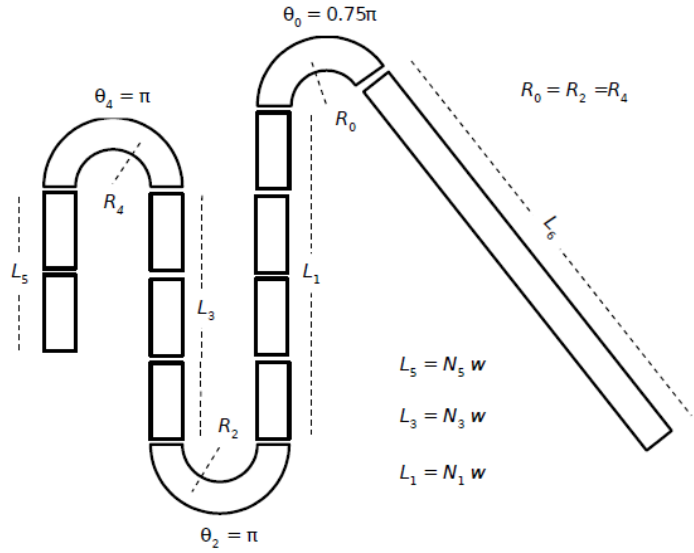

Figure 2. Model of self-siphon with length variation in fifth, third, and first segment indicates by set of number, $N_5$, $N_3$, and $N_1$, which in this case $N_5 = 2$, $N_3 = 3$, and $N_1 = 4$.

Friction of the head with the tube is defined as $f$, which is

$$f = -8\pi\eta\Delta h v, \tag{1}$$

as modified from Poiseuille equation in [20]. Equation (1) holds for both type of segments. For the vertical straight segments the force equations are

$$0 = ma_x \tag{2}$$

and

$$-mg + \rho g(y_w - y)A - f = ma_y. \tag{3}$$

And for the semi-circular segments

$$-mg\sin\theta R + \rho g(y_w - y)AR - fR = I\alpha, \tag{4}$$

with

$$I = mR^2, \tag{5}$$

$$\omega = \frac{v}{R}, \tag{6}$$

$$m = \Delta h \rho A, \tag{7}$$

Equation for time evolution of motion parameters are as follow

$$v_i(t + \Delta t) = v_i(t) + a_i\Delta t, \quad i = x, y, \tag{8}$$

$$r_i(t + \Delta t) = r_i(t) + v_i\Delta t, \quad i = x, y, \tag{9}$$

$$\omega(t + \Delta t) = \omega(t) + \alpha\Delta t, \tag{10}$$

$$\theta(t + \Delta t) = \theta(t) + \omega\Delta t, \tag{11}$$

Based on Equation (2) in the vertical segment the water does not flow in $x$-direction. In order to show whether the results in solving Equation (3) and (4) with MD simulation through Equation (8) – (11), there is a need to plot $r_x$, $r_y$, $v_x$, $v_y$, and $t$. It means that such relations

$$x = x_0 + R\cos\theta, \tag{12}$$

$$y = y_0 + R\sin\theta, \tag{13}$$

are needed, for the semi-circular segments. Parameters $x_0$ and $y_0$ are center of the semi-circular segments.

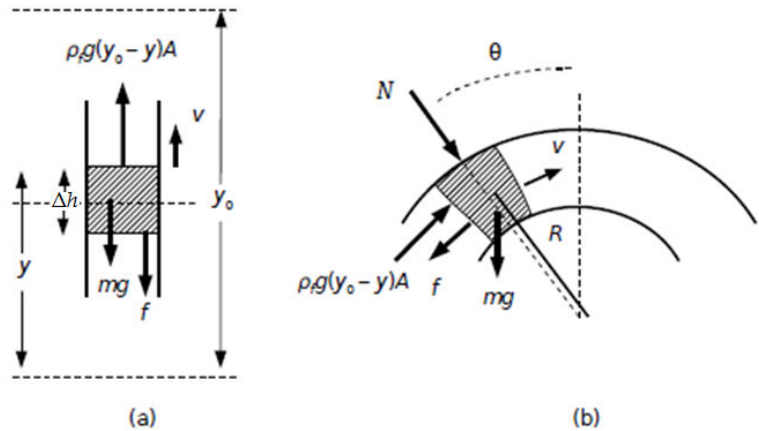

Figure 3. Diagram of considered forces applied to water element with thickness $\Delta h$: (a) in a vertical straight segment and (b) in circular segment.

The simulations are performed until the flow reaches the outlet from the inlet or when there is no flow, it must be ended when the flow changes its direction in a current segment, where is flowing.

## 3  Results and discussion

Configuration of $(N_5, N_3, N_1) = (4, 1, 4)$ is chosen to see whether the flow can follow the trajectory of self-siphon as it proposed in [17] using parametric equations. The result of position of the head as function of time is shown in Figure 4 and also the trajectory of self-siphon.

For this 414 configuration the fifth segment is passed between 0 s and 0.000928 s, the fourth segment is passed between 0.000928 s and 0.0010867 s, the third segment is passed between 0.0010867 s and 0.0014116 s, the second segment is passed between 0.0014116 s and 0.0015913 s, the first segment is passed between 0.0015913 s and 0.0021186 s, the zeroth segment is passed between 0.0021186 s and 0.0022143 s, and the sixth segment is passed between 0.0022143 s and 0.002792 s. It means the for 414 configuration $\tau = 0.002792$ s.

And for the velocity evolution in time, it is illustrated in Figure 5. In the chart of $v_y - v_x$, the direction of the head can be seen for every segment, more clearer as it seen in the chart of $r_y - r_x$. These results are obtained by using parameters value: $\Delta h = 0.01$, $\Delta t = 10^{-7}$, $\eta = 10^{-8}$. Variation of $N_5$, $N_3$, $N_1$ is conducted and 125 variations are noted, each from 0 to 4 for simulations as shown in Figure 6. The influece of $N_1$, $N_3$, $N_5$, independently to $\tau$ is also simulated and presented in Figure 7. It can be seen that as $N_3$ increasing the value of $\tau$ is also increasing, but as $N_1$ increasing value of $\tau$ is decreasing, this both trend do not match previously reported result [17]. Value of $\tau$ from simulations compared to previous result in [17] is 500 times lower as it should be. It means that the MD simulations need a scaling process or the implementation of Equation (1) needs more refinement since it depends on value of $\Delta h$ which is adjustable. Precise value of $\Delta h$ has not yet been investigated.

A height of inlet measured from water surface can be defined as

$$y_{\text{inlet}} = y_w - (N_1 - N_3 + N_5)w, \tag{13}$$

and then from Figure 6 it can seen that, especially from Figure 6 (e) that the flow can occur when
$$0.5N_1 + 1 - N_3 > 0, \ N_5 = 4. \tag{14}$$
Current results [18] show other condition than it is suggested by Equation (14) for the flow to occur. Using Equation (14) into Equation (13) gives about
$$y_{\text{inlet}} < y_w - 3w, \tag{15}$$
which is the condition that the self-siphon can flow the water. A critical height than can be defined from Equation (15) as
$$y_{\text{critical}} = y_w - 3w. \tag{16}$$

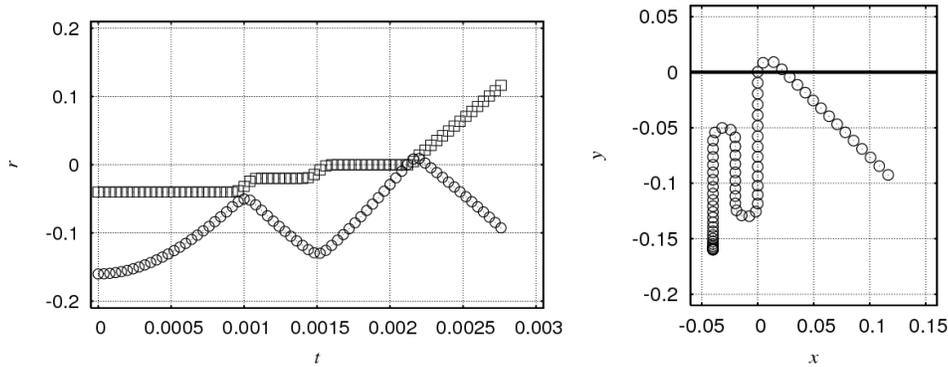

Figure 4. Position of water element for configuration of $(N_5, N_3, N_1)$ = (4, 1, 4): $r$ - t with symbols □ and ○ indicate $x$ and $y$, respectively (left) and $y$ - $x$ (right).

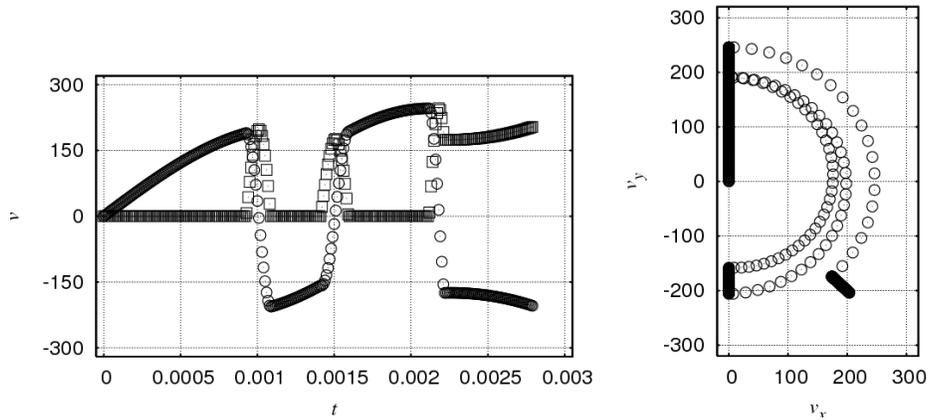

Figure 5. Velocity of water element (in m/s) for configuration of $(N_5, N_3, N_1)$ = (4, 1, 4): $v - t$ with symbols □ and ○ indicate $v_x$ and $v_y$, respectively (left) and (b) $v_y$ - $v_x$ (right).

Equation (16) is a requirement for the self-siphon to be able to flow the water. Height of inlet measured from water surface must be lower than this value.

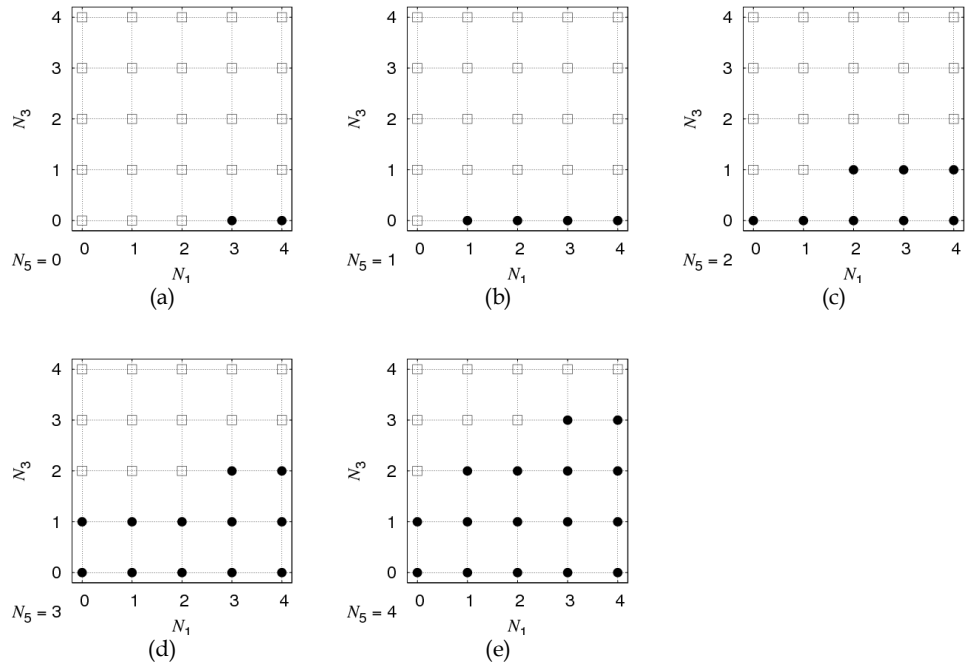

Figure 6. Variation parameter of $N_3$ and $N_1$ for different value of N5: (a) 0, (b) 1, (c) 2, (d) 3, and (e) 4. The symbols □ and ● indicate no flow and flow occurs, respectively, as predicted by MD simulations.

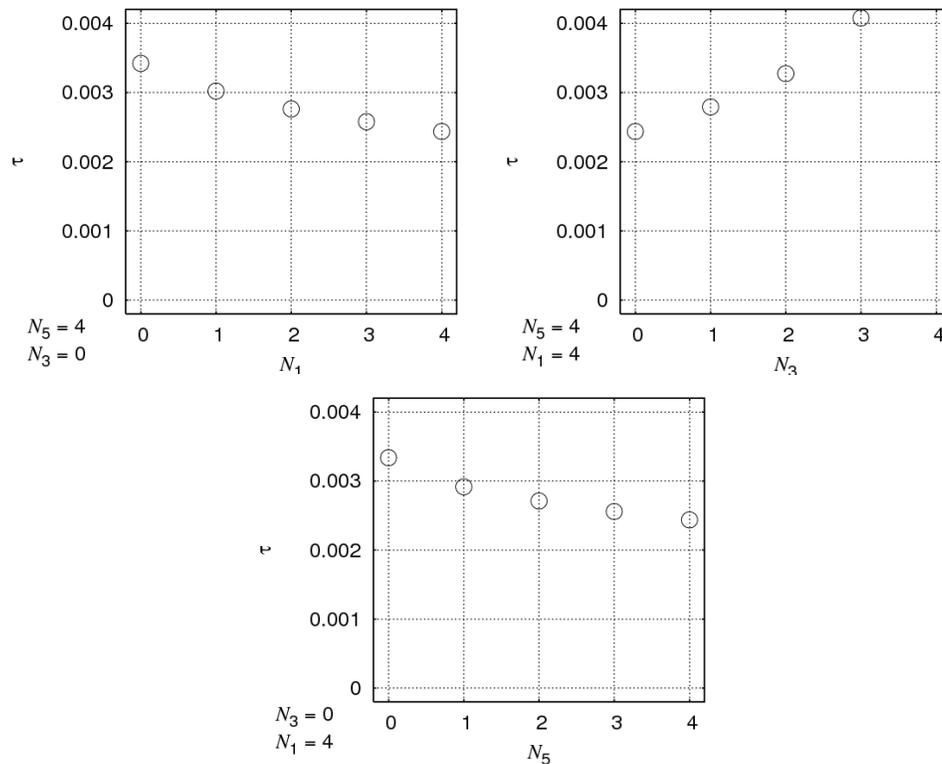

Figure 7. Dependency of time needed for water to flow from inlet to outlet of the self-siphon, τ (in s), to number of element in certain straight segment: $N_1$ (upper left), $N_3$ (upper right), and $N_5$ (lower center).

## 4 Conclusions

An model for simulating self-siphon flow using MD has been built. From the results it can be concluded that when water or fluid can flow, τ is increasing as $N_5$ and $N_1$ decreasing, but it is increasing as $N_3$ increasing. The last two results, on which τ dependent is, shows inverse behavior as previously reported in experiment. A critical height is also defined as the maximum value of inlet height, where the self-siphon still can flow the water.

## Acknowledgment

Authors would like to thank International Conference Grant IMHERE ITB and cooperation between Kanazawa University and Institut Teknologi Bandung in 2011 fur supporting presentation of this work, cooperation between Institut Teknologi Bandung and Ministerium of Religion Affair of Republic of Indonesia in 2009-2011 for supporting experiment part cited by this work, and Institut Teknologi Bandung Alumni Association Research Grant in 2010 for supporting simulation part of this work.